\documentclass[letterpaper]{article}
\usepackage{aaai18}
\usepackage{times}
\usepackage{helvet}
\usepackage{courier}
\usepackage{url}
\usepackage{graphicx}
\usepackage{array}
\usepackage{multirow}
\title{Evaluating Older Users' Experiences with Commercial Dialogue Systems: Implications for Future Design and Development}
\author{Libby Ferland \textsuperscript{1} \and Thomas Huffstutler \textsuperscript{1} \and Jacob Rice \textsuperscript{2} \and Joan Zheng \textsuperscript{1}\\ {\bf \Large \, \and Shi Ni \textsuperscript{1} \and Maria Gini \textsuperscript{1}}\\
\textsuperscript{1} Department of Computer Science and Engineering, University of Minnesota, Minneapolis, MN 55455\\
\textsuperscript{2} Wayzata High School, 4955 Peony Ln N, Plymouth, MN 55446 \\
}

\begin{document}
\maketitle
\begin{abstract}
Understanding the needs of a variety of distinct user groups is vital in designing effective, desirable dialogue systems that will be adopted by the largest possible segment of the population.  Despite the increasing popularity of dialogue systems in both mobile and home formats, user studies remain relatively infrequent and often sample a segment of the user population that is not representative of the needs of the potential user population as a whole.  This is especially the case for users who may be more reluctant adopters, such as older adults.  

In this paper we discuss the results of a recent user study performed over a large population of age 50 and over adults in the Midwestern United States that have experience using a variety of commercial dialogue systems.  We show the common preferences, use cases, and feature gaps identified by older adult users in interacting with these systems.  Based on these results, we propose a new, robust user modeling framework that addresses common issues facing older adult users, which can then be generalized to the wider user population.
\end{abstract}

%Joan & Thomas
\section{Introduction}
In the last decade, use of spoken dialogue systems (SDSs) has been rapidly increasing in popularity. Voice-controlled smart home devices, such as Amazon Echo devices and Google Home are currently being used in millions of households. Dialogue agents are also seeing increasingly heavy use as part of smart phone operating systems, with agents like Apple's Siri and Genie on Android devices leading the charge.  Despite their popularity, most commercial SDSs are primarily designed with healthy younger users in mind, overlooking older adult users and other large segments of the user population. This shortfall in design will become more readily apparent when considering future population trends. By 2050, over half of the US population is predicted to be over the age of fifty \cite{ortman2014}, and many of these adults which will likely interact with smart home devices and other dialogue systems on a regular basis. Furthermore, with this expected large increase in the size of the aged population, greater demands for care services will be made on already taxed healthcare systems.  Dialogue systems, especially smart home technologies, have been proposed as excellent resources to help alleviate strain on care providers.  However, despite this apparent need and demand, there remains a lack of current user studies that focus on older adult users' interactions with smart devices and dialogue systems and what they imply for the needs of this population.   
%smart home technologies are currently being developed that aim to provide support to this population within their own homes. Despite this apparent demand, there remains a lack of current user studies that focus on older adult users' interactions with smart devices and dialogue systems. 

In this study, we surveyed adults over age fifty on their experiences with commercially available dialogue systems.  We asked users to describe their interactions with the systems, including current use cases, failures and shortcomings, and any feature gaps they perceive and would like addressed.  The results of this study provide substantive, quantitative evidence that support many of the current focuses and research directions in the design and development of dialogue systems.  The results also reveal some very specific features, namely integrated, unimodal help features, that are most desirable to older users and may help serve as guideposts for directing the focus of future research.  %what users currently use the devices for and what tasks they wish to accomplish using SDSs. Being able to effectively identify the current gaps in the technology will hopefully enable us to create more effective aids for those who are unfamiliar with smart devices and open up use of these devices to a larger range of applications. 

In this paper, we first present some of the motivating and related work in the design and development of dialogue systems for a variety of users.  Next, we present key results from our study revealing important user behaviors and perceived and actual feature gaps.  We further discuss the implications of some of these results in terms of key research questions for the field, and propose that the need for context-sensitive, voice-activated help functions for these systems reveals a need to further focus research on developing robust user models and better predictive capabilities for user behavior.  Last, we detail some of the current and future work we are doing in applying deep learning techniques to larger bodies of user interaction data to develop a stronger user model.
%In this work, we design our survey to identify the gaps mentioned above as well as to provide some direction foe the smart device industry in the future. This study highlights the shortcoming of our current technology in the field as well as provide insights and suggestions into changes that will help improve these devices and increase their effectiveness to a growing user base.

%Libby
\section{Related Work}
Motivation for this work is drawn from general problems in the field of dialogue agents in addition to problems specific to older users.  The main motivation for undergoing this study stems from the desire to identify areas of acute interest in the development of dialogue systems as assistive technology, especially as support systems for %older and 
elderly users at  home.  

\subsection{Identifying Areas of Improvement} % for Conversational Agents}
Much literature has been devoted to identifying broad areas of interest in the development of conversational agents and assistive technology, independent of demographic-specific considerations.  In general, conversational agents have to face three different types of problems, including completing tasks, answering questions, and interacting socially \cite{Gao2018}.  These are extremely broad problems, but more and more approaches to creating effective dialogue systems acknowledge that they cannot be treated as wholly independent, and a complete end-to-end system must be expected to address all the different subtasks that make up the whole \cite{Gao2018} \cite{Rao2018}.  

It is specially important for assistive technologies to adopt this holistic view of task completion for dialogue systems, which requires a deeper examination of some  potential agent architectures and computational problems involved.  The design of intelligent personal assistants in particular provides some insight into potential areas of focus.  Milhorat et al. identify several key focus areas in the development of conversational agents as personal assistants \cite{Milhorat2014}.  These include: 1. Tracking longer dialogue histories; 2. Improving context awareness (ideally solely through dialogue, without involving additional sensor data); 3. Dynamic real-time adaptation to user behavior; and 4. Improved theoretical frameworks for problems such as task handling and task hierarchies.  Determining which of these areas is of acute interest to users is a different problem altogether, and one which we hope to address with the results of this study.

\subsection{Older Users Interacting with SDS}
A number of small-scale studies of older users interacting with smart technology help to narrow the list of areas of general improvement to those problems that are of particular importance to an older user demographic \cite{Schlogl2014}.  Previous studies show that in general, older users do interact with dialogue systems in a manner that is quite distinct from other groups.  In particular, older users tend to interact with dialogue agents in a manner much more consistent with interactions with living social agents, as opposed to younger users who are often much more task-driven \cite{Georgila2008} \cite{Moller2008}.  The most important aspects of interaction style in attempting to isolate key areas of improvement are that older users in general tend to participate in much more turn-taking in conversation, and use a much richer, more varied vocabulary with a greater number of social cue words. 

There are also mechanical features of speech for older people that may further dictate important areas of focus in the improvement of dialogue agents.  Older speakers often exhibit a greater number of speech disfluencies that can cause issues with successfully completing tasks using {SDS}s \cite{Bortfeld2001}.  First, older speakers use conversational repair strategies such as backtracking and error correction more frequently than do younger speakers, which dialogue systems may not handle gracefully.  A greater mechanical source of frustration, however, may be that older speakers tend to exhibit more frequent and longer pauses in speech.  These may cause issues with timeouts or reprompts, depending on the setup of a given dialogue system.

\subsection{Design and Acceptance with Older Users}
There are a number of different elements of user acceptance for older adults using assistive technology, especially technology with a social or intelligent component.  These include, but are not limited to trust, perceived adaptability, perceived usefulness, and perceived ease of use, among others \cite{Heerink2010}.  These dimensions are not often fully assessed for older users when designing new assistive technology.  A further, perhaps greater barrier to acceptance and adoption among older users of new technology is the failure to include representative users in design and testing phases.  This has been shown in a variety of studies, but is especially evident %when it comes to 
in the development of assistive technology \cite{Heerink2010} \cite{Dodd2017}.  

%Shi and Thomas
\section{Experimental Design}
\subsection{Setting}
This study was conducted in the D2D (Driven to Discover) facility at the Minnesota State Fair. The facility is designed for researchers to access a larger and more diverse pool of participants from fairgoers that is a  representative sample of the state’s population.  The majority of participants recruited in this setting were native English speakers and residents of Minnesota and Wisconsin.

\subsection{Participants}
%A display in front of the research booth gave the title of the study as well as the requirements for participants to be involved in the study. 

The participants that were actively recruited were adults of both sexes aged 50 or older, as well as those with severe vision impairment, as defined by commonly accepted occupational and legal standards \cite{Scheiman2007}.  Participants of any age with severe sensory impairment were included both as a point of comparison with an even wider age range, and as a representative sample to help pinpoint any issues that may specifically pertain to users of dialogue systems with the types of sensory impairments commonly seen in an older population. 
%meeting the criteria as described previously. 

\subsection{Procedure}
Potential participants were given an introduction to the background and objectives of the study. Those who wished to participate were asked to sign a consent form were shown to a table where they were given either an iPad or a paper copy to complete the survey based upon their preference.

The most significant portion of the data collected %from the questionnaires 
was with regards to the participants' experiences with smart devices (i.e., Alexa, Siri, Google Home, etc.) as well as their thoughts on current shortcomings of smart devices and potential improvements that could be made. Participants were first asked about their experience with smart devices in terms of frequency. Based on this response, if they identified themselves as users of voice activated technology either currently or in the past, they were directed to section one. Section one asked participants about which specific smart devices they used the most, the application of the smart devices they used, and most thoroughly, their thoughts on the voice interaction experience of using the smart devices. The interest in the users experience with voice interactions of smart devices was with regards to the difficulty of the interaction, their preferences for voice interaction, and their thoughts on the current constraints of the technology. 
Section two of the survey focused on participants who initially responded that they were non-users of smart technology. Questions in that section asked about experiences with smart devices they have used, reasons for their limited use of smart technologies,  impressions on the voice capabilities of smart devices, and tasks for which they would like to %be able 
to use smart devices.

Questions from both sections of the survey were designed to follow the {A}lmere model for assessing acceptance and acceptability of assistive technology in older adults, as proposed by Heerink et al. \cite{Heerink2010} The {A}lmere model identifies `intention to use' as a heavy contributing factor in whether or not older users actually adopt assistive technology.  Intention to use in turn is a combination of user perceptions, including perceptions of enjoyment, usefulness, ease of use, and trust and trustworthiness.  The questions in this study are designed specifically to develop a more detailed picture of how well commonly available dialogue systems address these factors, and to further identify major acceptability gaps.  These questions were also specifically chosen in the hopes of developing a more detailed picture of the connections between any identified acceptability gaps and the current state-of-the-art and relevant research questions in the field.

\section{Results}
The survey instrument in this study contained two sections, one pertaining to experiences with `smart technology' in general, and the second pertaining to experiences with voice interaction specifically.  Questions in both sections were fairly similar as a form of quality control and verification of consistency in participant responses.  For the sake of brevity and maintaining focus on the most relevant results for this work, this paper will only discuss the responses to the second, voice specific, section.

All free response questions had two independent raters.  All questions were optional, so response rate may also vary per question.  The population size for each question is given.  Most free response questions allowed multiple answers per participant.

%*******N MUST BE PRESENTED SUBDIVIDED INTO DEVICE TYPES*******
%*********DON'T FORGET TO ADD***************
\subsection{Demographics}
The total number of respondents was N = 174.  Of these, 145 participants indicated at least some experience with dialogue systems.  36 participants were primarily users of smart home systems, 102 were users of mobile devices, and 7 were not otherwise categorized.  Table 1 contains a summary of the primary device type and frequency of use for respondents.  
%frequency of use and device type
\begin{table}[h!]
\begin{tabular}{|b{1.75cm}||b{1cm}|b{1cm}|b{1cm}|b{1cm}|}  
 \hline
 %\multirow{2}{2em}{}
  & \multicolumn{4}{|c|}{Frequency of Use}\\\cline{2-5} Device Type & Daily & Weekly & Monthly & Rarely\\\cline{2-5}

% \multirow{3}{5em}{Primary Device Type}& \multicolumn{4}{|c|}{Frequency of Use}\\

%\multirow{2}{5em}{}%  & Daily & Weekly & Monthly & Rarely\\\cline{2-5}
\hline
\hline
 Smart Home &  0.43 & 0.27 & 0.17 & 0.13 \\%13  & 8 & 5 & 4 \\
 Smart Phone & 0.35 & 0.29 & 0.23 & 0.13 \\%33 & 27 & 22 & 12\\
 Other  & 0.09 & 0.09 & 0.00 & 0.82 \\ %& 2 & 2 & 0 & 2\\
 \hline
 \hline
 Total & 0.33 & 0.25 & 0.18 & 0.23 \\%& 48 & 37 & 27 & 18 \\
 \hline
\end{tabular}
\caption{\textit{Device type and usage frequency among participants (N = 126.  N = 30 smart home device users, 94 smart phone users)}}
\label{Table 1: }
\end{table}
\subsection{User Preferences Over Interaction Style}
%social preferences
Participants were asked to identify their preferred interaction style with dialogue agents.  Responses for this question were hard-coded and covered a variety of social, personally adaptable, or impersonal interaction styles.  Responses are once again grouped by primary device type.
\begin{table}[h!]
\begin{tabular}{|b{1.3cm}||b{1cm}|b{0.7cm}|b{0.55cm}|b{0.55cm}|b{0.8cm}| b{0.75cm}|}
\hline
& \multicolumn{6}{c|}{Interaction Preference} \\\cline{2-7}

Device Type & A, $\lnot S$ & A, S & S & I & All A & All S \\
\hline
\hline
Smart Home & 0.33 & 0.36 & 0.06 & 0.25 & 0.69 & 0.42 \\
Smart Phone & 0.45 & 0.23 & 0.13 & 0.20 & 0.68 & 0.35 \\
Other & 0.33 & 0.17 & 0.33 & 0.17 & 0.50 & 0.50 \\
\hline

    \end{tabular}
    \caption{\textit{Dialogue system interaction style preferences by primary device type.  Options abbreviated for length.  A = adapted to user interaction style and history, S = social or conversational agent, I = impersonal agent, no preference over adaptability to user.  N = 144 respondents. N = 36 smart home users, 102 mobile device users. }}
    \label{Table 2}
\end{table}

Among smart home device users, 69\% preferred an agent with some ability to adapt to user interaction styles and history.  42\% preferred a social or relational agent.  25\% of users had a preference for strictly impersonal interactions.  Among smart phone or other mobile device users, 68\% preferred an adaptable agent, 35\% preferred a social or relational agent, and 20\% of users preferred strictly impersonal interactions.  There is no significant difference in any of the preferences between the different device types.

\subsection{Frequent Use Cases}
Participants were  asked to identify their most common or favorite uses of voice interaction with their primary device.  
\begin{table}[h!]
    \centering
    \begin{tabular}{|b{0.85cm}||b{0.5cm}|b{0.55cm}|b{0.55cm}|c|c|c|b{0.5cm}|}
    \hline
    & \multicolumn{7}{|c|}{Interaction Type} \\\cline{2-8} 
    Device Type & Ph. & Dict.& Mus. & Gen. & Struct. & O & NS \\
    \hline
    \hline
    Smart Home & 0.03 & 0.09 & 0.11 & 0.51 & 0.20 & 0.06 & 0.00 \\%1 & 3 & 4 & 18 & 7 & 2 & 0 \\
    Smart Phone & 0.16 & 0.24 & 0.03 & 0.24 & 0.19 & 0.11 & 0.03 \\%15 & 23  & 3 & 23 & 18 & 10 & 3 \\
    Other & 0.00 & 0.00 & 0.00 & 0.50 & 0.50 &  0.00 & 0.00 \\
    \hline
    \end{tabular}
    \caption{\textit{Most frequent uses for older users of dialogue systems.  Ph. = phone, Dict. = text dictation, Mus. = music, Gen. = general information (unstructured queries), Struct. = structured queries (weather, time, etc.), O = other, NS = not sure.  N = 134.  N = 35 smart home users, 95 mobile device users}}
    \label{tab:my_label}
\end{table}

We differentiate %in coding 
between structured and unstructured queries.  Structured queries are highly transactional interactions that have %tend towards 
shallow dialogue trees (i.e., asking for weather r directions, querying for time).  Unstructured queries are interactions to find more general information with a high likelihood of %necessity of or desire for
follow-up questions (i.e., web searches).  Surprisingly, despite frequent frustrations identified elsewhere in the survey, users most often interact with dialogue systems %in order 
to obtain general information.
\subsection{Frequent Difficulties}
We next asked participants to identify their most common sources of difficulty when interacting with dialogue systems.  

\begin{table}[h!]
    \centering
    \begin{tabular}{|b{0.85cm}||b{0.5cm}|c|b{0.5cm}|b{0.5cm}| c |b{0.5cm}|c|}
    \hline
    & \multicolumn{7}{|c|}{Source of Difficulty} \\\cline{2-8} 
    Device Type & GQ & Dict. & UL &SR & U(NS) & O & NS \\
    \hline
    \hline
    Smart Home & 0.29 & 0.03 & 0.03 & 0.03 &  0.23 & 0.26 & 0.13 \\%9 & 1 & 1 & 1 &  7& 8 & 4 \\
    Smart Phone & 0.18 & 0.10 & 0.07 & 0.07 & 0.18 & 0.30 & 0.09 \\%17 & 10 & 7 & 7 & 17 & 29 & 9 \\
    Other & 0.14 & 0.14 & 0.00 & 0.00 & 0.29 & 0.14 & 0.29 \\%1 & 1 & 0 & 0 & 2 & 1 & 2 \\
    \hline
    \end{tabular}
    \caption{\textit{Most frequent difficulties experienced by users in voice interactions with dialogue systems.  GQ = general/complex questions, Dict. = text dictation, UL = unusual language or names, SR = speech recognition, U(NS) = understanding (non-specific), O = other, NS = not sure.  N = 134.  N = 31 smart home users, 96 mobile device users.}}
    \label{tab:my_label}
\end{table}

Most difficulties identified in this question relate to issues with `understanding.'  In this work, we are interested in examining differences between experiences with language understanding (i.e., intent recognition, {NLU}, {NLG}) and speech recognition.  For this and all other free response questions, the coding scheme presented here is strict, and responses on `understanding' that do not differentiate between sound and language are categorized as understanding (non-specific).

As might be expected from frequent use cases, the most commonly identified source of difficulty with {SDS}s is in unstructured queries and (non-specific) understanding.  The indicated difficulties do not vary significantly between smart home and mobile technologies when proportions are taken into account, except for the case of text dictation.

\subsection{Desired Features and Improvements}
Participants were asked to identify any perceived feature gaps in the commercial {SDS} they use most often.  This question had the lowest rate of response out of the entire study, and many responses were highly specific and thus categorized as `other.'
Most of the commonly identified desired features are %in fact 
existing features in the {SDS}s most commonly seen in this study.  Interestingly, respondents in this question did not specifically indicate speech recognition, unlike in every other question.  Instead participants gravitated towards issues of {NLU} and {NLG}.

\begin{table}[h!]
    \centering
        \begin{tabular}{|b{0.85cm}||b{0.5cm}|c|b{0.5cm}|b{0.5cm}| c |b{0.5cm}|c|}
    %\hline
    %\begin{tabular}{|b{1cm}||c|c|c|c|c|c|c|}
    \hline
    & \multicolumn{7}{|c|}{Desired Features} \\\cline{2-8} 
    Device Type & CS & UH& MD & LU & UNS & O & NS \\
    \hline
    \hline
    Smart Home & 0.14 & 0.11 & 0.14 & 0.04 & 0.07 & 0.25 & 0.25 \\
    % 4 & 3 & 4 & 1 & 2 & 7 & 7\\
    Smart Phone & 0.04 & 0.08 & 0.20 & 0.04 & 0.04 & 0.23 & 0.37 \\
    % 3 & 6 & 15 & 3 & 3 & 18  & 28\\
    Other & 0.00 & 0.50 & 0.25 & 0.00 & 0.00 & 0.00 & 0.25 \\
    % 0 & 2 & 1 & 0 & 0 & 0 & 1\\
    \hline
    \end{tabular}
    \caption{\textit{Commonly desired features and improvements for commercial dialogue systems.  CS = context switching, UH = user help functions, MD = multiple device integration, LU = language understanding, UNS = understanding (non-specific), O = other, NS = not sure.  N = 108. N = 28 smart home users, 76 mobile device users}}
    \label{tab:my_label}
\end{table}

\subsection{Attractive Qualities of Commercially Available {SDS}}
The last two questions of the study asked participants to discuss features, issues, or impressions that most attract or repel them when using {SDS} and interacting with intelligent agents via voice.

\begin{table}[h!]
    \centering
    \begin{tabular}{|b{0.85cm}||c|c|c|c|c|c|}
    \hline
    & \multicolumn{6}{|c|}{Attractive Aspects} \\\cline{2-7} 
    Device Type & Speed & EoU & Conven. & HF & O & NS \\
    \hline
    \hline
    Smart Home & 0.40 & 0.33 & 0.06 & 0.15 & 0.06 & 0.03 \\
    %13 & 11 & 2 & 5 & 2 & 1 \\
    Smart Phone & 0.34 & 0.25 & 0.06 & 0.26 & 0.17 &0.01 \\
    %32 & 24 & 6 & 25 & 16 & 1 \\
    Other & 0.60 & 0.60 & 0.00 & 0.00 & 0.20 & 0.00 \\
    \hline
    \end{tabular}
    \caption{\textit{Most attractive elements of {SDS} and voice activated technology among older users.  EoU = ease of use, Conven. = convenience, HF = hands-free, O = other, NS = not sure.  N = 133. N = 33 smart home users, 95 mobile device users.  This question was multiple response.}}
    \label{tab:my_label}
\end{table}

Most responses to this question were fairly generic and did not vary widely.  It is worth noting, however, that despite clearly indicated frustrations and issues with fundamental elements of {SDS} technology, like speech recognition and language understanding, many users still view the technology as easy to use and convenient.

\subsection{Undesirable Qualities of Commercially Available {SDS}}
The last question in the survey echoed earlier, more general questions about smart technology and asked participants to identify their biggest source of frustration or least desirable feature in using {SDS}s.

\begin{table}[h!]
    \centering
    \begin{tabular}{|b{0.85cm}||c|c|c|b{0.5cm}|b{0.5cm}|b{0.5cm}|c|}
    \hline
    & \multicolumn{7}{|c|}{Desired Features} \\\cline{2-8} 
    Device Type & SR & LU & UNS & SP & LC & O & NS \\
    \hline
    \hline
    Smart Home & 0.07 & 0.17 & 0.20 & 0.03 & 0.13 & 0.37  & 0.07 \\
    %2 & 5 & 6 & 1 & 4 & 11 & 2\\
    Smart Phone & 0.21 & 0.12 & 0.23 & 0.04 & 0.04 & 0.28 & 0.09 \\
    %19 & 11 & 21 & 4 & 4 & 25  & 8\\
    Other & 0.00 & 0.17 & 0.00 & 0.33 & 0.33 & 0.00 & 0.33 \\%0 & 1 & 0 & 2 & 2 & 0 & 2\\
    \hline
    \end{tabular}
    \caption{\textit{Commonly identified sources of frustration and undesirable aspects of using {SDS} in home and mobile systems.  SR = speech recognition, LU = language understanding, UNS = understanding (non-specific), SP = security and privacy, LC = learning curve/knowledge gap, O = other, NS = not sure.  N = 126.  N = 30 smart home users, 90 mobile device users.  Multiple responses possible.}}
    \label{tab:my_label}
\end{table}

Specificity in responses to this question varied wildly ("Not understanding me" vs. "Siri does not understand who the Packers are"), but follow the same general themes as other questions.  Poor understanding of all types was by far the most commonly identified undesirable quality of {SDS}.

% Rao needs to go in here..somewhere (possibly also future work)
\section{Discussion}
This study focused  on older users, who as a user group are generally less targeted for study and perhaps more hesitant to adopt new technology.  However, they are also the fastest growing segment of the population worldwide, and a number of studies and applications indicate that an aging population may derive some of the greatest benefit from artificial assistants and other technological support in a variety of ways, including memory support and maintaining independence.  This user group, therefore, cannot be overlooked and should be viewed as a great source for information about what a general user population truly wants out of dialogue systems.

While some of the findings of this study may echo previous work, the strength of this study is two-fold.  First, the population size examined here is many times larger than most similar studies, which often boast population sizes of 30 or less.  Second, the fact that this study focuses on a less well-examined portion of the user population provides insight into what features of dialogue systems are truly generalizable, and how developers and researchers can focus efforts to create systems tailored to the needs of relatively untapped potential user groups.

The absolute conclusion of this study is that user modeling and user intent recognition are perhaps the two most important problems underlying many of the issues and desires of the wider user population.  Time and again, responses to questions indicate a desire for features and abilities that are underpinned by the need for more robust capabilities in these two areas.  Interaction style preferences, perception of feature gaps, and common user frustrations all point to the fact that an agent that is better capable of predicting user actions, and tailoring its behavior to match the behavior of individual users, should be the most immediate goal in the development of further dialogue systems.

\subsection{Implications of User Preferences Over Interaction Style}
A natural conversational relational agent is of course the end goal in developing dialogue systems and indeed almost any artificial agent.  However, in this study users indicate fairly strongly that it is not the social ability of dialogue agents that interest them so much as the ability of agents to adapt to them.  In fact, a significant number of users have a distinct preference for sterile agents that are not at all social.  This indicates that more immediate efforts in research and development should be directed towards problems like user intent recognition, context switching, and the development of a robust, adaptable user model that captures unique patterns in user behavior revealed by spoken interactions and transaction histories.

\subsection{Implications of Frequent Use Cases}
One of the more surprising results of this study was that despite any perceived or actual capability gaps in dialogue systems, users have already embraced them as participants in less-structured vocal interactions.  By far the most commonly indicated use for dialogue agents on both smart home and mobile platforms is in answering general or free-form queries such as might appear in a web search.  The inclusion of general queries as one of the biggest sources of user frustration further supports that this is the use case users really want these systems to be sufficient for.  The obvious conclusion from this result is that natural language understanding should remain one of the most important areas of research in developing dialogue systems.  However, this also supports the conclusion that developing agents with better predictive capabilities based on a deeper understanding of user behavior and use patterns should be a major focus in assistive agent research.

\subsection{Implications of Common Frustrations and Perceived Feature Gaps}
The root of many user frustrations is naturally in difficulties with both speech recognition and natural language understanding.  However, a number of user frustrations and perceived gaps in agent capability actually stem from a lack of awareness of what these agents can actually do.  A number of responses to this survey also indicated that the accessibility of educational materials is a barrier in both learning to use and being willing to learn to use these systems.  

%cite Hawkey, etc. here
Part of the onus in addressing these issues certainly lies with the manufacturers and developers of commercial systems.  However, the results of this study obliquely suggest that one of the best ways to address issues with the availability of educational materials is to keep user education as unimodal as possible.  This is especially helpful in the case of elderly or otherwise impaired users who may have sensory deficits that act as barriers to accessing more traditional media avenues for user materials like tutorials and help menus.  A number of studies already exist that indicate that dialogue systems are ideal for training exercises in a variety of applications and for all age groups, and may be especially helpful for older users in terms of cognitive support and skill maintenance \cite{Dodd2017} \cite{Greenaway2013}.  Literature in supporting adults with memory issues further paints intelligent assistants as ideal helpmates in situations where questions may be repetitive and users may need frequent reminders and guidance in procedural tasks \cite{Hawkey2005} \cite{Konig2016}, such as might be the case with na\"ive users learning to use a new piece of technology.

%
%\subsection{}
%Libby
%leave this part for WPE maybe
%\section{Proposed User Model}

%Libby
\section{Conclusions and Future Work}
In this work we present key results of a very large user study with older users of dialogue systems.  The responses from study participants shows that while many wants, needs, and current use cases for older users mirror those of other user groups, older users do have some unique needs and desires, namely in learning to use new and continuously evolving systems.  These results indicate that older users and other na\"ive users of {SDS} technology may   benefit from more closely integrated system help and guidance, which means that problems like  robust user modeling and intent prediction are vital in addressing the needs of a wider user population.

Based on the results of this study, we will focus on the development of a robust user model based on pattern detection and extraction from extended user interaction histories.  To do this, we will be collecting user interaction data from na\"ive users of dialogue systems interacting with an intelligent personal assistant.  After developing this dataset, we will use deep learning techniques to find features of voice interactions, including user intents and context tracking, that are unique to individual users and can be used as the basis for a generalized user model that can be learned and improved by dialogue agents with each successive user interaction.

%Libby
\section{Acknowledgements}
The authors would like to thank Charlotte Marien for assistance in data collection.  The authors also gratefully acknowledge the support of an {NSF} {NRT} training grant, ``Graduate Training Program in Sensory Science: Optimizing the Information Available for Mind and Brain'' (Grant DGE--1734815) and the {ROSE-HUB} {I/UCRC} in undertaking this study.
%to add or not to add - possible that a complete analysis should be published separately since it needs to be more detailed (and there's a ton of metanalysis to be done)

%A report containing a complete analysis of all survey data, including metanalyses, is available from the authors upon request.

\bibliography{oldersdsbib}

\begin{thebibliography}{}

\bibitem[\protect\citeauthoryear{Bortfeld \bgroup et al\mbox.\egroup
  }{2001}]{Bortfeld2001}
Bortfeld, H.; Leon, S.~D.; Bloom, J.~E.; Schober, M.~F.; and Brennan, S.~E.
\newblock 2001.
\newblock Disfluency rates in conversation: Effects of age, relationship,
  topic, role, and gender.
\newblock {\em Language and speech} 44(2):123--147.

\bibitem[\protect\citeauthoryear{Dodd, Athauda, and Adam}{2017}]{Dodd2017}
Dodd, C.; Athauda, R.; and Adam, M. T.~P.
\newblock 2017.
\newblock {Designing User Interfaces for the Elderly: A Systematic Literature
  Review}.
\newblock {\em Australasian Conference on Information Systems}  1--11.

\bibitem[\protect\citeauthoryear{Gao, Galley, and Li}{2018}]{Gao2018}
Gao, J.; Galley, M.; and Li, L.
\newblock 2018.
\newblock Neural approaches to conversational ai.
\newblock {\em arXiv preprint arXiv:1809.08267}.

\bibitem[\protect\citeauthoryear{Georgila \bgroup et al\mbox.\egroup
  }{2008}]{Georgila2008}
Georgila, K.; Wolters, M.; Karaiskos, V.; Kronenthal, M.; Logie, R.; Mayo, N.;
  Moore, J.~D.; and Watson, M.
\newblock 2008.
\newblock A fully annotated corpus for studying the effect of cognitive ageing
  on users ' interactions with spoken dialogue systems.
\newblock {\em Proceedings of the 6th International Conference on Language
  Resources and Evaluation (LREC)} (1):938--944.

\bibitem[\protect\citeauthoryear{Greenaway, Duncan, and
  Smith}{2013}]{Greenaway2013}
Greenaway, M.~C.; Duncan, N.~L.; and Smith, G.
\newblock 2013.
\newblock The memory support system for mild cognitive impairment: Randomized
  trial of a cognitive rehabilitation intervention.
\newblock {\em International Journal of Geriatric Psychiatry} 28(4):402--409.

\bibitem[\protect\citeauthoryear{Hawkey \bgroup et al\mbox.\egroup
  }{2005}]{Hawkey2005}
Hawkey, K.; Inkpen, K.~M.; Rockwood, K.; McAllister, M.; and Slonim, J.
\newblock 2005.
\newblock Requirements gathering with alzheimer's patients and caregivers.
\newblock In {\em Proc. 7th Int'l ACM SIGACCESS Conference on Computers and
  Accessibility},  142--149.

\bibitem[\protect\citeauthoryear{Heerink \bgroup et al\mbox.\egroup
  }{2010}]{Heerink2010}
Heerink, M.; Kr\"{o}se, B.; Evers, V.; and Wielinga, B.
\newblock 2010.
\newblock Assessing acceptance of assistive social agent technology by older
  adults: the almere model.
\newblock 2:361--275.

\bibitem[\protect\citeauthoryear{K\"{o}nig \bgroup et al\mbox.\egroup
  }{2016}]{Konig2016}
K\"{o}nig, A.; Malhotra, A.; Hoey, J.; and Francis, L.~E.
\newblock 2016.
\newblock Designing personalized prompts for a virtual assistant to support
  elderly care home residents.
\newblock In {\em Affective Interaction with Virtual Assistants within the
  Healthcare Context},  278--282.

\bibitem[\protect\citeauthoryear{Milhorat \bgroup et al\mbox.\egroup
  }{2014}]{Milhorat2014}
Milhorat, P.; Schlogl, S.; Chollet, G.; Boudy, J.; Esposito, A.; and Pelosi, G.
\newblock 2014.
\newblock Building the next generation of personal digital assistants.
\newblock In {\em Advanced Technologies for Signal and Image Processing
  (ATSIP), 2014 1st International Conference on},  458--463.

\bibitem[\protect\citeauthoryear{M{\"{o}}ller, G{\"{o}}dde, and
  Wolters}{2008}]{Moller2008}
M{\"{o}}ller, S.; G{\"{o}}dde, F.; and Wolters, M.~K.
\newblock 2008.
\newblock A corpus analysis of spoken smart-home interactions with older users.
\newblock {\em LREC}  735--740.

\bibitem[\protect\citeauthoryear{Ortman \bgroup et al\mbox.\egroup
  }{2014}]{ortman2014}
Ortman, J.~M.; Velkoff, V.~A.; Hogan, H.; et~al.
\newblock 2014.
\newblock {\em An aging nation: the older population in the United States}.
\newblock US Census Bureau, Economics and Statistics Administration, US
  Department of Commerce.

\bibitem[\protect\citeauthoryear{Rao, Ture, and Lin}{2018}]{Rao2018}
Rao, J.; Ture, F.; and Lin, J.
\newblock 2018.
\newblock Multi-task learning with neural networks for voice query
  understanding on an entertainment platform.
\newblock In {\em Proceedings of the 24th ACM SIGKDD International Conference
  on Knowledge Discovery \& Data Mining},  636--645.
\newblock ACM.

\bibitem[\protect\citeauthoryear{Scheiman, Scheiman, and
  Whittaker}{2007}]{Scheiman2007}
Scheiman, M.; Scheiman, M.; and Whittaker, S.
\newblock 2007.
\newblock {\em Low vision rehabilitation: A practical guide for occupational
  therapists}.

\bibitem[\protect\citeauthoryear{Schl{\"{o}}gl, Garschall, and
  Tscheligi}{2014}]{Schlogl2014}
Schl{\"{o}}gl, S.; Garschall, M.; and Tscheligi, M.
\newblock 2014.
\newblock {Designing Natural Language User Interfaces with Elderly Users}.
\newblock {\em Workshop on Designing Speech and Language Interactions at the
  ACM SIGCHI Conference on Human Factors in Computing Systems}  1--4.

\end{thebibliography}
\bibliographystyle{aaai}
\end{document}